\documentclass[amsmath,aps,prb,superscriptaddress,showpacs,amssymb,secnumarabic,eqsecnum,twocolumn,floatfix]{revtex4-1}
\usepackage{graphicx}
\usepackage{pstricks}
\usepackage{hyperref,url}

\newrgbcolor{blanco}{255 255 255}
\newcommand{\ba}{\begin{eqnarray}}
\newcommand{\ea}{\end{eqnarray}}
\def\be{\begin{equation}}
\def\ee{\end{equation}}
\def\n{\nonumber\\}

\def\vk{\mathbf{k}}

\begin{document}
\title{Fermi surface instabilities at finite Temperature}
%\title{Some exact results for the nematic to isotropic transition at finite temperature}
%\title{Some exact results for the phase diagram of interacting Fermi systems in the presence of Pomeranchuk instabilities
%at finite temperature
%}
%Fermi surface instabilities for systems with factorizable interactions
%or separable interactions
\author{P. Rodr\'\i guez Ponte}
\affiliation{Instituto de F\'\i sica de La Plata and Departamento de F\'isica, Universidad Nacional de La Plata, C.C. 67, 1900 La Plata, Argentina}
\author{D.C. Cabra}
\affiliation{Instituto de F\'\i sica de La Plata and Departamento de F\'isica, Universidad Nacional de La Plata, C.C. 67, 1900 La Plata, Argentina}
\author{N. Grandi}
\affiliation{Instituto de F\'\i sica de La Plata and Departamento de F\'isica, Universidad Nacional de La Plata, C.C. 67, 1900 La Plata, Argentina}
\affiliation{Abdus Salam International Centre for Theoretical Physics, Associate Scheme \\ Strada Costiera 11, 34151, Trieste, Italy}

\begin{abstract}
We present a new method to detect Fermi surface instabilities for interacting systems at finite temperature. We first apply it to a list of cases studied previously, recovering already known results in a very economic way, and obtaining most of the information on the phase diagram analytically. As an example, in the continuum limit we obtain the critical temperature as an implicit function of the magnetic field and the chemical potential $T_c(\mu,h)$. By applying the method to a model proposed to describe reentrant behavior in $Sr_3Ru_2O_7$, we reproduce the phase diagram obtained experimentally and show the presence of a non-Fermi Liquid region at temperatures above the nematic phase.
\end{abstract}


\maketitle

\section{Introduction}
%
%FL picture and Pomeranchuk instabilities
%
Landau's theory of the Fermi liquid (FL)\cite{Legget,Baym} is the standard description of weakly interacting fermionic systems. The basic assumption is that, at the microscopic level, there is a statistical ensemble of elementary excitations that behave as fermionic quasiparticles with well defined momenta and energies. These quasiparticles are assumed to have long lifetimes, and their distribution function shows a step at finite momentum defining the Fermi surface (FS) in momentum space.
In this picture, low energy excitations are described by infinitesimal deformations of the FS. When one of such distortions leads to a reduction in the free energy, we have a breakdown of the FL description due to what is known as a Pomeranchuk instability \cite{Pomeranchuk}. This kind of instabilities has been intensively studied during the last fifteen years. In particular, broken symmetry phases arise due to electron-electron interactions, within the Landau-Fermi liquid description \cite{Metzner0,Metzner01,Gros,Wu,Wuotro,Bonfim,Quintanilla3,Jaku}. In some cases, such instabilities lead to nematic phases
\cite{nematic1,*nematic1a,*nematic1b,*nematic1c,yamase,Metzner_PRL,nematic2,Hankevych,Metzner1,Wu1,nematic_MF,YK,Lawler,Lawlerb,Kee1,nematic3,Metzner2,Metzner2b,Quintanilla2,Hinkov,Hinkovb,Arovas}.
A comprehensive review about nematic Fermi fluids with an extensive list of references could be found in Ref. [\onlinecite{nematics}]. \\

%autobombo
In previous works we developed and applied a general method to detect Pomeranchuk instabilities. With it, we were able to obtain the instability regions for several systems. While the first work \cite{nos1} was mainly advocated to study cases involving simple interactions (as $s$-wave and $d$-wave interactions) and zero-temperature square lattice systems, the method was extended to incorporate temperature dependence (mainly within a low temperature expansion approximation) and flavor degrees of freedom \cite{nos3,nos4}. Also, application to systems with different lattices, as the honeycomb case has been successfully accomplished \cite{nos2}.
%NICOLAS agregue lo de abajo para que se entienda el parrafo que viene despues
The approach consisted in decomposing the deformations of the FS in a set of basis functions or ``channels'', and then studying the contribution of each such channel to the free energy.

%nuevo metodo
It is the aim of this work to present a more efficient method to diagnose FS instabilities. With this new approach, by making use of the symmetries of the interaction function, we are able to find the FS instabilities in a much more economic way, as compared with previous approaches. We can now obtain the unstable regions avoiding much of the time-consuming numerical calculations that were faced previously. Furthermore, with this method there is no need of any extrapolation to higher channels to determine the phase boundaries, which in many cases can be obtained analytically. In addition, the present method allows to study finite temperature problems without the need of a low temperature expansion.
The method is discussed in detail in Section \ref{sec:method}.\\

%resultados del nuevo metodo
In Section \ref{sec:examples} we apply this method to several cases for both zero and finite temperature. There we show that applying the present method to the previously studied cases, the reported results are recovered. Moreover, for much of the cases discussed in this section, we get the instability condition as an analytic expression of the parameters involved. For instance, when we study the stability of a finite temperature system in an external magnetic field with a dispersion relation corresponding to the continuum limit, we are able to find an implicit relation for the critical temperature as a function of the applied magnetic field and the chemical potential.\\

%aplicacion a lo de yamase
In Section \ref{sec:rutenato} we consider a $d$-wave interaction function. This kind of interaction has been proposed to describe nematic Fermi liquids \cite{nematics}  and shown to be relevant in the discussion of the reentrant behavior manifested in ultra-clean samples of $Sr_3Ru_2O_7$ at low temperatures \cite{grigeraotrosa,*grigeraotrosb,*grigeraotrosc,grigera2011}. There, as the external magnetic field is increased, the compound goes from an isotropic Fermi liquid phase to a nematic one, and then back to the isotropic phase. The intermediate phase forms a dome-shaped region in the $T$ {\em vs.} $h$ phase diagram. A review of the physical properties of $Sr_3Ru_2O_7$ can be found in [\onlinecite{grigera2012}].

By studying this model within a mean field approximation, Yamase and co-workers \cite{yamase2005} obtained - for some values of the parameter's space - a wide region of stability for the isotropic Fermi liquid and the characteristic dome-shaped region for the nematic Fermi liquid. They also obtained the observed behavior for the phase transitions between both phases (continuous transitions for the upper temperature part of the dome and discontinuous for the lower temperature part). Since then, it has been well established \cite{yamaseresto,*yamaserestob,*yamaserestoc} that this approach is a good starting point in the construction of a phenomenological description of the observed behavior in $Sr_3Ru_2O_7$.

A major problem of the mean-field approach, when used alone, is that it fails to describe the non Fermi liquid region of the phase diagram: the model seems to indicate that the isotropic Fermi liquid phase is also present there, even for higher temperatures, above the dome-shaped region. Our goal is to improve the mean-field approach, by combining it with the Pomeranchuk stability analysis developed in the present paper, to obtain a phase diagram schematically shown in Fig.1.\\

\begin{figure}
\includegraphics[width=0.45\textwidth]{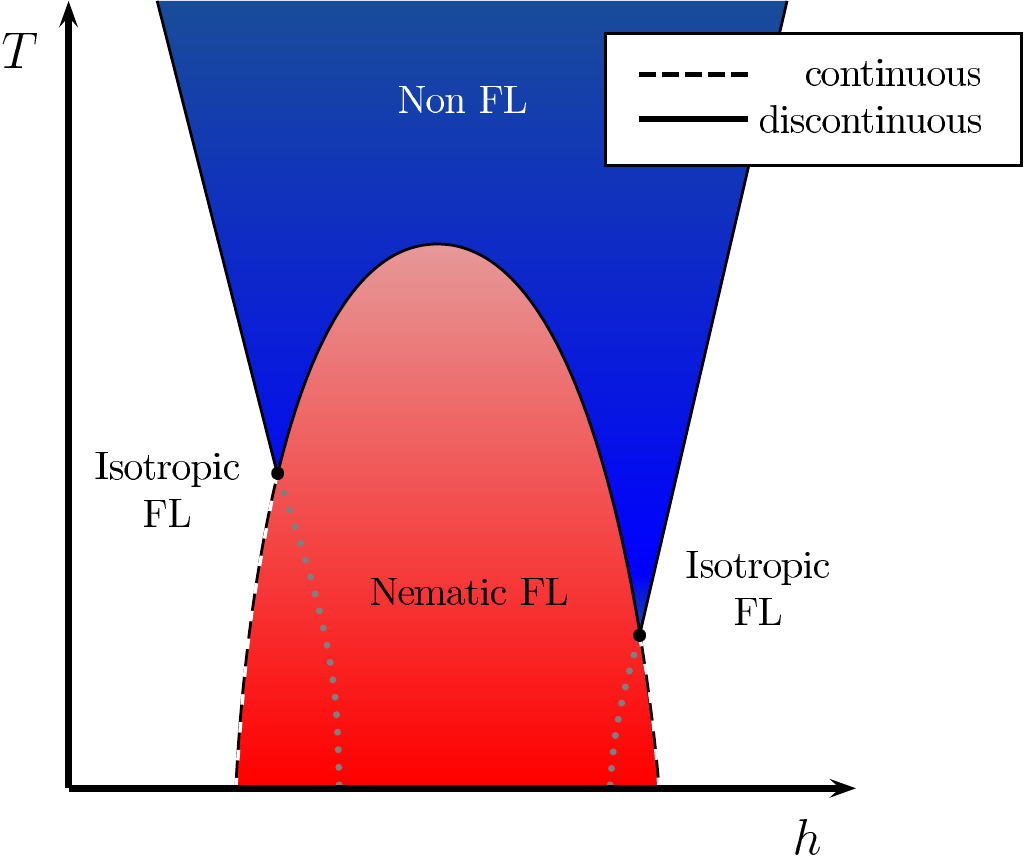}
\caption{(Color online) Schematic phase diagram, showing both Fermi liquid (FL) and non-Fermi liquid phases. The continuous (dashed) lines correspond to continuous (discontinuous) phase transitions. Gray dotted lines correspond to continuous transitions between isotropic and nematic FL phases that arise within our stability analysis but which are shown to be preempted by a discontinuous transition. See Section  \ref{sec:rutenato} for details.}
\label{fig:schematicdiagram}
\end{figure}
%We conclude this work with some final general considerations, in
Section \ref{sec:conclusions} contains the conclusions.

\section{The method}
\label{sec:method}
According to Landau's theory of the Fermi liquid, the change in the Landau free-energy $\Omega=E-TS-\mu^\alpha N_\alpha$ can be written as a functional of the change $\delta n^\alpha(\vk)$ in the equilibrium distribution function at finite chemical potential $\mu^\alpha$ for each particle species $\alpha\in[1\dots N_f]$. To first order in the interaction. It reads
\begin{widetext}
\ba &&
\delta \Omega= \sum_\alpha\int d^{2}\!\vk\,(\varepsilon^\alpha(\vk)\!-\!\mu^\alpha)\,\delta n_\alpha(\vk) +  \frac{1}{2}\sum_\alpha\sum_\beta\int \!d^{2}\!\vk \!\int\!d^{2}\!\vk'
f^{\alpha\beta}(\vk,\vk')\,\delta n_\alpha(\vk)\delta n_\beta({\vk'}) .\ \
\label{eq:deltaE}
\ea
\end{widetext}
Here $\varepsilon^\alpha(\vk)$ is the dispersion relation that controls the free dynamics of the system, and the interaction function $f^{\alpha\beta}(\vk,\vk')$ can be related to the low energy limit of the two particle vertex. We will decompose the interaction function as
\be
f^{\alpha\beta}(\vk,\vk')= \sum_{ij}^N U^{ij}d_i^\alpha(\vk)d_j^\beta(\vk')
\label{eq:decomp}
\ee
where $\{d_i^\alpha\}_{i \in I}$ represents a given basis of the space of functions on momentum space.
%NICOLAS modifique la frase en negrita
For most of the interactions used in the literature, such basis can always be chosen in such a way that the sum has a finite number of terms $N$.

With this at hand, we will prove that the variation of the Landau free energy can be rewritten as
\ba
\delta\Omega&=&\sum_{ij}^{N}c^ic^jM_{ij}
\label{eq:stability_matrix}
\ea
where $\{c^i\}_{i\in [1,..,N]}$ are arbitrary real numbers parameterizing the deformations of the occupation numbers $\delta n^\alpha(\vk)$, and we have defined the ``stability matrix'' $M_{ij}$ in the form
\be
M_{ij} =(d_i|d_j)
+\sum_{kl}^N
U^{kl}
\langle d_k|d_i\rangle
\langle d_l|d_j\rangle
\label{eq:stability_matrix_2}
\ee
in terms of the positive-definite bilinear forms
\ba
(\psi|\phi)&=&
\int d^2k\sum_{\alpha}
(\varepsilon^\alpha({\bf k})\!\!-\!\mu^\alpha) F''[\mu^\alpha\!\!-\!\varepsilon^\alpha({\bf k})]\,
\psi^\alpha({\bf k})\phi^\alpha({\bf k})
\nonumber\\
\langle\psi | \phi\rangle&=&\int d^2k
\sum_{\alpha}
F'[\mu^\alpha\!\!-\!\varepsilon^\alpha({\bf k})]
\,
\psi^\alpha({\bf k})
\phi^\alpha({\bf k}).
\label{eq:forms}
\ea
The key idea in Pomeranchuk's analysis is to characterize
deformations $\delta n^\alpha(\vk)$ that would lead to $\delta \Omega<0$, then pointing to an instability of the system.  Then, according to (\ref{eq:stability_matrix}), in order to have a positive energy for all sets of $\{c^i\}_{i\in[1,..,N]}$, the matrix $M_{ij}$ has to be positive-definite. A simple way to check positivity is to verify that all the minors of the matrix $M_{ij}$ are positive, {\em i.e.}
\be
\mbox{Min}\left\{\phantom{\frac12}\!\!\!\!\mbox{Minor}_i(M)\right\}_{i\in[1,N]}>0
\label{eq:cond}
\ee
This is our central result: to find the Pomeranchuk instabilities of a given system, one has to compute the corresponding stability matrix (\ref{eq:stability_matrix_2}), evaluate its minors which are functions of $\mu^\alpha$, $T$ and $U_{ij}$, and find the region in phase space in which they become negative.

It has to be pointed out that the calculation of the stability matrix involves only a finite number $2N^2$ of integrals, which in most cases are straightforward to compute, and in some of the problems studied in the literature can be even found analytically.

\subsubsection*{Proof of the result}
In order to prove the result presented above, we insert the decomposition of the interaction function (\ref{eq:decomp}) into the variation of the Landau free energy (\ref{eq:deltaE}), to get
\begin{widetext}
\ba
\delta \Omega&=& \sum_\alpha\int \!d^{2}\!\vk\,(\varepsilon^\alpha(\vk)\!-\!\mu^\alpha)\,\delta n_\alpha(\vk)
+  \frac{1}{2}\sum_{ij}^NU^{ij}
\left(
\sum_\alpha\int \!d^{2}\!\vk\, d_i^\alpha(\vk)\delta n_\alpha(\vk)
\right)
\left(\sum_\beta\int\!d^{2}\!\vk'd_j^\beta(\vk')\delta n_\beta({\vk'})
\right)
\,.\ \
\label{eq:deltaE2}
\ea
\end{widetext}
The next step is to
parameterize any given deformation of the occupation numbers with the help of a new function $\delta^\alpha({\bf k})$, as
\ba
\delta n^\alpha({\bf k})\!&=&\!F[\mu^\alpha\!\!-\!\varepsilon^\alpha({\bf k}) +\delta^\alpha({\bf k})]-F[\mu^\alpha\!\!-\!\varepsilon^\alpha({\bf k}) ]=
\n
\!&\simeq&\! F'[\mu^\alpha\!\!-\!\varepsilon^\alpha({\bf k})]\delta^\alpha({\bf k})\!+\!\frac12 F''[\mu^\alpha\!\!-\!\varepsilon^\alpha({\bf k})] {\delta^\alpha({\bf k})}^2%+\cdots
\,,\n
\label{eq:deltaN}
\ea
where $F[x]$ is the Fermi distribution
\be
F[x]=\frac1{e^{-x/{k_BT}}+1}
\ee
and a prime $(')$ denotes a derivative with respect to $x$.

With the help of eqs. (\ref{eq:deltaN}), we can rewrite the variation in $\delta \Omega$ to lowest order in $\delta^\alpha({\bf k})$ as
\be
\delta\Omega=
\frac12\left( (\delta |\delta )
+
\sum_{ij}^NU^{ij}
\langle d_i|\delta \rangle
\langle d_j|\delta \rangle
\right)
\label{eq:energy_bilinears}\ee
where the bilinear forms were defined in (\ref{eq:forms}). If we now decompose $\delta^\alpha$ in the same basis $\{d_i^\alpha\}_{i \in I}$ as the interaction function, namely
\be
\delta^\alpha({\bf k})=\sum_i^N c^i d_i^\alpha({\bf k}) + \delta_\perp^\alpha({\bf k})
\ee
it is easy to prove that we can always choose $\delta_{\perp}^\alpha({\bf k})$ as satisfying
\be
\forall_i: \ \ \ \langle \delta_\perp|d_i\rangle=0
\ee
The proof is as follows: given the basis $\{d_i^\alpha\}_{i \in I}$ we separate it into two parts $\{d_i^\alpha\}_{i \in [1,\dots,N]}\cup \{d_i^\alpha\}_{i \in (I-[1,\dots,N])}$ and replace the second factor by another set of basis functions $\{\tilde d_i^\alpha\}_{i \in (I-[1,\dots,N])}$ to which we have subtracted the part non-orthogonal to $\{d_i^\alpha\}_{i \in [1,\dots,N]}$, namely $\tilde d_i^\alpha(\vk) = d_i^\alpha(\vk) -\sum_{j\in [1,\dots,N]} \langle d_i|d_j\rangle \langle d_j|d_j\rangle^{-1}  d_j^\alpha(\vk)$.
By a similar argument, we can further decompose
\be
\delta^\alpha_\perp({\bf k})=\sum_i^N \tilde c^i d_i^\alpha({\bf k}) + \delta_{\perp\perp}^\alpha({\bf k})
\ee

~

\noindent where $\delta_{\perp\perp}^\alpha({\bf k})$ is defined as satisfying
\be
\forall_i: \ \ \ (\delta_{\perp\perp}|d_i)=0
\ee
Recalling the previous condition in $\delta_{\perp}({\bf k})$ we get
\be
\sum_i^N \tilde c^i\langle d_i
|d_j\rangle
+\langle \delta_{\perp\perp}
|d_j\rangle
=0
\label{eq:c_tildes}
\ee
which implies that the $\tilde c^i$ are not independent parameters, but they are given in terms of the $c^i$'s.
Replacing into the expression (\ref{eq:energy_bilinears}) for energy we obtain
\begin{widetext}
\ba
\delta\Omega&=&\frac12\sum_{ij}^N\left(c^ic^j
\left((d_i|d_j)
+\sum_{kl}^N
U^{kl}
\langle d_k|d_i\rangle
\langle d_l|d_j\rangle
\right)
%+\n&&
+2 c^i \tilde c^j (d_i|d_j)
+\tilde c^i \tilde c^j (d_i|d_j)
\right)
+\frac12\, (\delta_{\perp\perp}|\delta_{\perp\perp})
\ea
\end{widetext}
By comparing the above expression with equation (\ref{eq:energy_bilinears}), we see that the last term represents the fluctuation of a free fermion with $U_{ij}=0$ when perturbed with $\delta_{\perp\perp}({\bf k})$. This implies that this last term is always positive and does not contribute with instabilities to the above sum. Then we can safely put $\delta_{\perp\perp}({\bf k})=0$. This in turn implies that, according to eq.(\ref{eq:c_tildes}) and if the matrix $\langle d_i |d_j\rangle$ is invertible, then the coefficients $\tilde{c}^i$ have to vanish. We then obtain
\ba
\delta\Omega&=&\frac12\sum_{ij}^N c^ic^j
\left((d_i|d_j)
+\sum_{kl}^N U^{kl}
\langle d_k|d_i\rangle
\langle d_l|d_j\rangle \right)\n
\label{eq:stability_matrix_proof}
\ea
In conclusion, the energy can be written as the quadratic form (\ref{eq:stability_matrix}), and we have proved our main result.
\section{Application to known examples}
\label{sec:examples}
In this section we will apply the method presented above to systems whose Pomeranchuk instabilities have been previously studied in the literature by alternative methods \cite{nos1,nos2,nos3,nos4,Metzner_PRL,Hankevych,Quintanilla2}. We will start by applying it to the zero temperature spinless case, and progress later to more complicated situations.
\subsection{Zero temperature}
In the zero temperature limit, both bilinear forms (\ref{eq:forms}) contains a Dirac $\delta$-function in the $\mu^\alpha\!\!-\!\epsilon({\bf k})$ variable. Indeed, when $T\to0$ we have  $F'[x]\to\delta(x)$ and $-x F''[x]\to-x\delta'(x)=\delta(x)$. This implies that the bilinear forms coincide, and depend only on the values of the integrand evaluated at the Fermi surface
\ba
\langle\psi|\phi\rangle&=&(\psi | \phi)=\sum_{\alpha}\int d^2k\,\delta(\mu^\alpha\!\!-\!\varepsilon^\alpha({\bf k})) \,\psi^\alpha({\bf k}) \phi^\alpha({\bf k})\n
\label{eq:formsT0}
\ea
This integral can be further simplified by parameterizing the Fermi surface of species $\alpha$ with the help of a parameter $s^\alpha\in S^1$ as ${\bf k}={\bf k}(s^\alpha)$. We get\cite{nos2}
\ba
\langle\psi|\phi\rangle&=&(\psi | \phi)=\sum_{\alpha}
\oint \frac{ds^\alpha|{\bf k}'(s^\alpha)|}{|\nabla \varepsilon^\alpha({\bf k}(s^\alpha))|}\,\psi^\alpha({\bf k}(s^\alpha)) \phi^\alpha({\bf k}(s^\alpha))\n
\label{eq:formsT02}
\ea
An immediate consequence of this is that all the $\tilde c^i$ excitations vanish.
The stability matrix reads
\ba
M_{ij} &=&\sum_k^N\left(\langle d_i|d_k\rangle\left(\delta_{kj}
+\sum_{l}^N
U^{kl}
\langle d_l|d_j\rangle
\right)\right)=
\n
&\equiv&\sum_k^N
\langle d_i|d_k\rangle M^{(o)}_{kj}
\ea
Notice that the factor $\langle d_i|d_k\rangle$ correspond to the stability matrix of a zero temperature Fermi system without interactions, {\em i.e.} a Fermi gas. Since such Fermi gas is obviously stable, then by the considerations of the previous section, its matrix $\langle d_i|d_k\rangle$ is positive definite. The condition of positivity of $M_{ij}$ is reduced to the positivity of the ``zero temperature stability matrix" $M^{(o)}_{kj}$. In other words, at zero temperature the Fermi liquid will be stable as long as
\be
\mbox{Min}\left\{\phantom{\frac12}\!\!\!\!\mbox{Minor}_i(M^{(o)})\right\}_{i\in[1,N]}>0
\label{eq:condT0}
\ee
In what follows we apply this condition to some simple cases
\subsubsection*{Spinless case}

At zero temperature and in the absence of spin, the index $\alpha$ is not present and the sums in the bilinear forms (\ref{eq:formsT0}) collapse to a single term. The resulting expression is particularly simple
\ba
\langle\psi|\phi\rangle&=&
\oint \frac{ds|{\bf k}'(s)|}{|\nabla \varepsilon({\bf k}(s))|}\,\psi({\bf k}(s)) \phi({\bf k}(s))
\label{eq:formsT01}
\ea

We will start by studying the simplest possible case of an interaction containing a single function $d(\bf k)$. In other words, we will chose an interaction function that can be written as
\be
f({\bf k},{\bf k}')=U d({\bf k})d({\bf k'})
\label{eq:separablespinless}
\ee
In this case the zero temperature stability matrix contains a single entry, and the stability condition (\ref{eq:condT0}) reads
\be
M^{(o)} =
1
+U
\langle d|d\rangle>0
\label{eq:condT0sep}
\ee
with $\langle d|d\rangle$ as in (\ref{eq:formsT01}).

As the first concrete example of the above simplifications, we can use the method to study the Fermi liquid in the square lattice that was previously investigated by a different method in [\onlinecite{nos1}]. In a square lattice, the Fermi surface is defined by
\be
\varepsilon({\bf k})=-2t\left(\cos k_x +\cos k_y \right)\equiv  \mu
\ee
where $t$ is the hopping parameter, and $\mu$ the chemical potential. It can be parameterized as
\ba
k_x&=&{\rm arccos}\left(-\frac \mu{4t}+s\right)
\n
k_y&=&{\rm arccos}\left(-\frac \mu{4t}-s\right)
\ea
where $-1+\frac {|\mu|}{4t}<s<1+\frac {|\mu|}{4t}$.
With this at hand, the bilinear form (\ref{eq:formsT01}) reads
\be
\langle\psi|\phi\rangle=\frac1{2t}\oint ds
\frac{\psi(s)\phi(s)}{\sqrt{s^4-2 \left(\left(\frac \mu{4t}\right)^2\!+\!1\right) s^2+\left(\left(\frac \mu{4t}\right)^2\!-\!1\right)^2}}
\label{eq:square}
\ee

\begin{figure}
\includegraphics[width=0.45\textwidth]{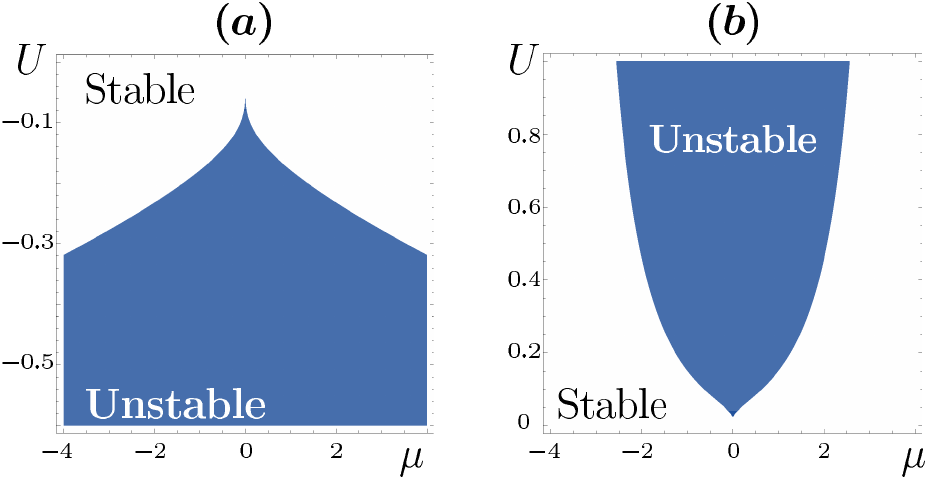}
\caption{Instability region for a Fermi liquid with an $s$-wave (a) and a $d$-wave (b) interaction in the square lattice. Note that both plots coincide with the extrapolation to higher channels of the plots presented in Ref.[\onlinecite{nos1}].}
\label{inestsydwave}
\end{figure}

Now in the case of an $s$-wave interaction in the square lattice, we have
\be
f({\bf k},{\bf k}')=U\,,
\label{eq:swave}
\ee
that can be decomposed in the form we need by using a single function $d(\vk)$ given by
\be
d({\bf k})=1
\label{eq:swaved}
\ee
In this case the integral (\ref{eq:square}) can be performed explicitly, an we get an analytic result for the stability condition $M^{(o)}>0$ in parameter space
\be
1+\frac{16 \,U}{4t+\mu}\, {\rm Re}\!\left[ K\!\left(\frac{(4t-\mu)^2}{(4t+\mu)^2}\right)\right]>0
\ee
where $K(\cdots)$ denotes the complete elliptic integral of the first kind. With this expression we immediately get the phase diagram of Fig.\ref{inestsydwave}(a). Note that the plot coincides remarkably with the extrapolation to higher channels of the plot presented in Ref.[\onlinecite{nos1}].

If instead we have a $d$-wave interaction, namely
\be
f({\bf k},{\bf k}')=-U(\cos k_x-\cos k_y)(\cos k'_x-\cos k'_y)
\ee
that can again be represented in terms of a single basis function $d({\bf k})$, given in this case by
\ba
d({\bf k})&=&\cos k_x -\cos k_y =2s
%=
%\n
%&=&2 \sec (2 s) \left(\frac{1}{2 J|_{g=0}}-1\right)
\ea
after performing the integral (\ref{eq:square}) to obtain $\langle d|d\rangle$ we get for the stability condition $M^{(o)}>0$
\small
\be
1-\frac{4 U}{t^2}(4t+\mu) {\rm Re}\!\left[\left(K\!\left(\frac{(4t-\mu)^2}{(4t+\mu)^2}\right)\!-\!E\!\left(\frac{(4t-\mu)^2}{(4t+\mu)^2}\right)\right)\right]
>0
\ee
\normalsize
where $E(\cdots)$ represents the complete elliptic integral of the second kind. The resulting phase diagram is shown in Fig.\ref{inestsydwave}(b). Again the diagram coincides remarkably with the extrapolation to higher channels of the plots presented in Ref.[\onlinecite{nos1}].
\vspace{1cm}

Now we can progress into more complicated situations. For example we can consider non-square lattices, and/or interactions with multiple components, in which case the zero temperature stability matrix would contain multiple entries.

As an example, in Ref.[\onlinecite{nos2}] a honeycomb lattice was studied. There, on site interactions with strength $u$ and nearest neighbor interactions with strength $v$ were considered.
In its factorized form, the interaction function reads
\be
f({\bf k},{\bf k}')=\sum_{i=1}^7U^{i}d_i({\bf k})d_i({\bf k'}),
\label{fgraphene}
\ee
where the coefficients are $U^{1}=(2 u + 3 v )/(4 \pi)^2$ and $U^{i}=-v/(4 \pi)^2$ for $i=2$ to $7$, and the functions $d_i({\bf k})$ are
\ba
d_1({\bf k})&=&1,\\
d_2({\bf k})&=&\cos{({\bf k.\delta_1}+\phi)}\,\,,\,\,d_3({\bf k})=\cos{({\bf k.\delta_2}+\phi)}\nonumber\\
d_4({\bf k})&=&\cos{({\bf k.\delta_3}+\phi)}\,\,,\,\,d_5({\bf k})=\sin{({\bf k. \delta_1}+\phi)}\nonumber\\
d_6({\bf k})&=&\sin{({\bf k.\delta_2}+\phi)}\,\,,\,\,d_7({\bf k})=\sin{({\bf k. \delta_3}+\phi)}\nonumber
\ea
with ${\bf\delta_1}=a(\frac{1}{2},\frac{\sqrt{3}}{2})$, ${\bf\delta_2}=a(\frac{1}{2},\frac{-\sqrt{3}}{2})$ and ${\bf\delta_3}=a(-1,0)$.

\begin{figure}
\includegraphics[width=0.45\textwidth]{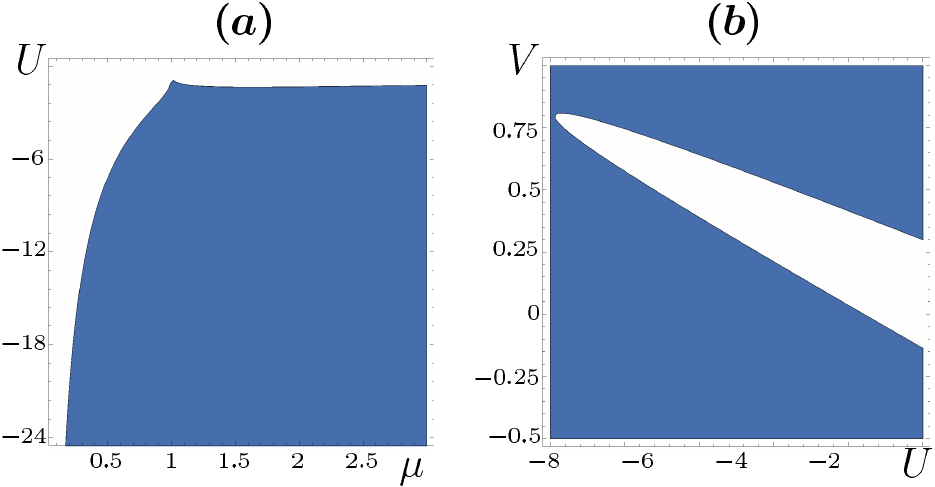}
%\begin{picture}(1,1)(0,0)
%  \put(-18,-8){{\small $\mu$}}
%  \put(-125,105){{\small $u$}}
%\end{picture}
\caption{Instability regions (shaded) for a Fermi liquid with the interactions given by (\ref{fgraphene}) in the honeycomb lattice. Planes $v=0$ (in fig. (a)) and   $\mu=0.95$ (fig. (b)). Both plots seem to recover the extrapolation to higher channels of what was reported Ref.[\onlinecite{nos2}].}
\label{graphene}
\end{figure}

Introducing this information into the expression of our zero temperature stability matrix, it becomes a $7\times7$ matrix given by
\be
M_{ij}^{(o)}=\delta_{ij} +U^{i} \langle d_i|d_j\rangle > 0.
\ee
Using it in (\ref{eq:condT0}) we get the phase diagram of Fig.\ref{graphene}. Again, notice the coincidence with the results obtained by different methods.

\subsubsection*{Spinful case in an external magnetic field}

A further step would be the inclusion of spin, which allows for the coupling to an external magnetic field. We will limit our study to the case of a simple interaction that can be written in terms of a single $d^\alpha(\vk)$ as
\be
f^{\alpha\beta}({\bf k},{\bf k}')=U d^\alpha({\bf k})d^\beta({\bf k'})
\label{eq:intspin3}
\ee
In the case of a spin antisymmetric interaction in the square lattice with nearest neighbor hopping, the simplest example of the kind studied in [\onlinecite{nos4}] is given by
\be
d^{\alpha}({\bf k})=\alpha%\cos(\theta_{\vk})
\label{eq:intspin}
\ee
with $\alpha=\pm1$.
With this interaction and recalling the results of the previous section for the square lattice, we immediately get from the stability condition $M^{(o)}>0$ the requirement
\be
\label{eq:coninciden}
1+\sum_\alpha\frac{16\, U}{4t\!+\!\mu\!+\!\alpha h}\,{\rm Re}\!\left[ K\!\left(\frac{(4t\!-\!\mu\!-\!\alpha h)^2}{(4t\!+\!\mu\!+\!\alpha h)^2}\right)\right]>0
\ee
where $2h$ is the applied magnetic field.

\begin{figure}
\includegraphics[width=0.45\textwidth]{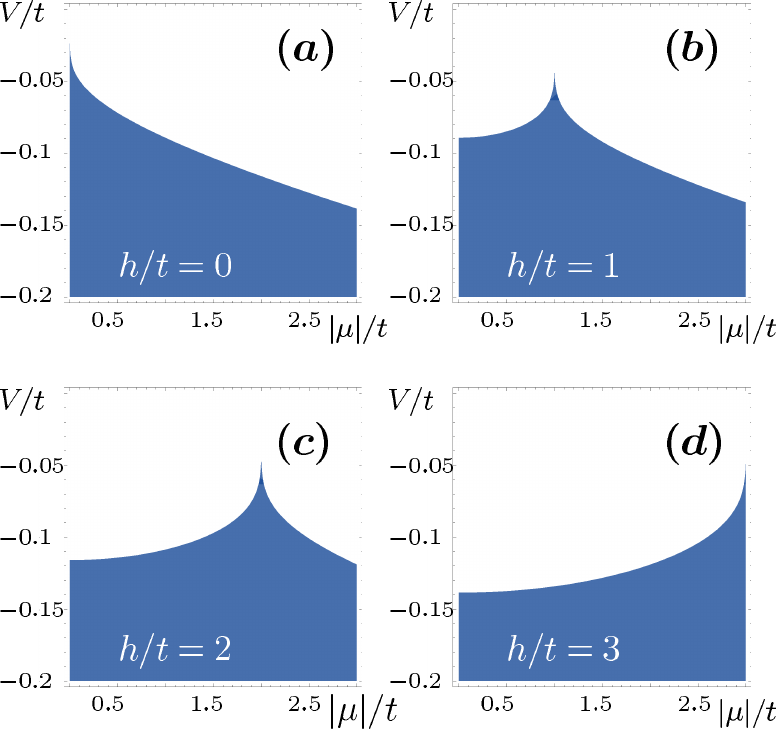}
\caption{(Color online) (a)-(d) Instability regions (shaded) for external magnetic field going from $ h/t=0$  to $ h/t=3$. The plots seem to recover the extrapolation to higher channels of what was reported Ref.[\onlinecite{nos4}].}
\label{fig:spinsplit}
\end{figure}

A somewhat more complicated interaction studied in [\onlinecite{nos4}] is the one given by
\be
d^{\alpha}({\bf k})=\alpha\, \frac {k_x}{|\vk|}
\label{eq:intspin2}
\ee
Remarkably, the zero temperature stability matrix for this interaction is given by the same formula (\ref{eq:coninciden}), with $U$ replaced by $U/2$. The result coincides with the spin-split instability found in [\onlinecite{nos4}], and in particular, as it can be seen in Fig.\ref{fig:spinsplit}, reproduces the re-entrant behavior reported there, that was observed in experiments.

\subsection{Finite temperature}
With the training of the previous sections, we can now proceed to study the finite temperature case. Now, the bilinear forms introduced in (\ref{eq:forms}) are no longer equal, and the stability matrix may be more involved.

\subsubsection*{Spinless case}
With a simple interaction in terms of a single $d(\vk)$ as in (\ref{eq:separablespinless}), we get the stability matrix as
\be
M =
(d|d)
+
U
\langle d|d\rangle^2>0
\ee

\begin{figure}
\includegraphics[width=0.45\textwidth]{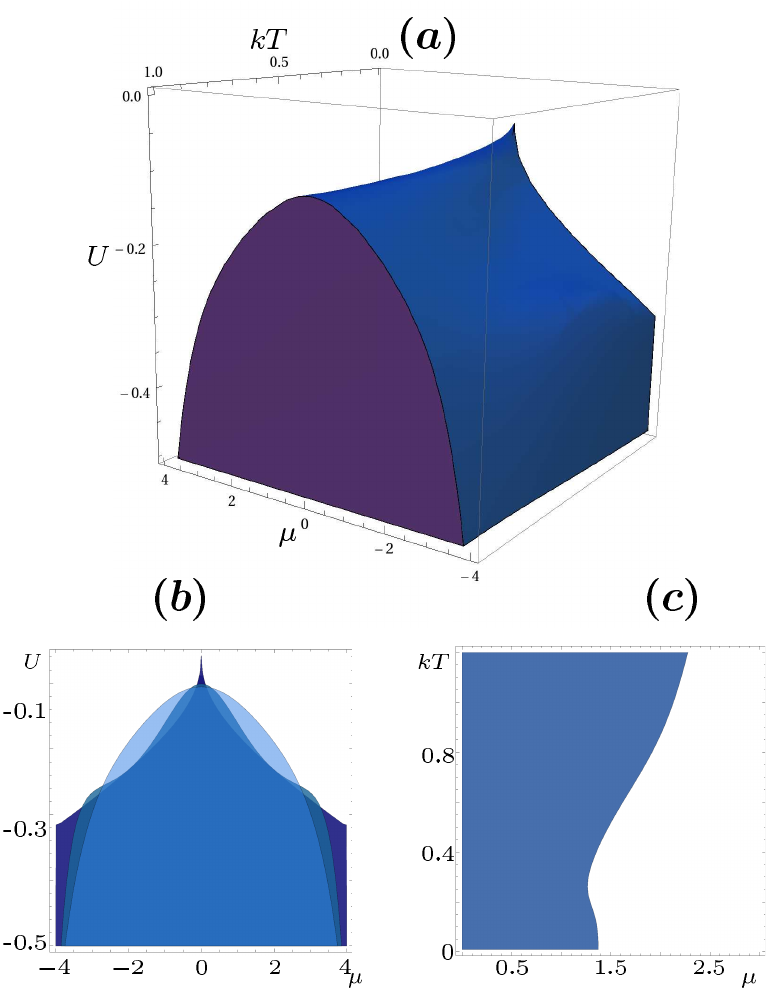}
\caption{(Color online) Instability regions (shaded) for s-wave interaction function. (a) 3D phase diagram showing unstable regions in the space of parameters $\mu$, $U$ and $T$. (b) Instability regions for $kT=0.01,0.5$ and $1$. (c) Instability regions for $U=-0.2$.}
\label{swaveT}
\end{figure}

In the case of an $s$-wave interaction (\ref{eq:swave})-(\ref{eq:swaved}), the integrals in the $\langle d|d\rangle$ and $(d|d)$ brackets
have to be evaluated numerically. We get for the phase diagram the one displayed in  Fig.\ref{swaveT}

Again, note that the stability region coincides with the previously published results obtained by a different method in Ref.[\onlinecite{nos3}]. In particular, in that work an expansion in powers of $kT/\mu$ was needed, while the present calculation is exact to any order in this parameter.
\subsubsection*{Spinful case in an external magnetic field}
To test our method on a spinful system at finite temperature, we will use the simple dispersion relation corresponding to the continuum limit, {\em i.e.}
\be
\varepsilon^\alpha(\vk)=\frac {1}{2m}\left(k^2_x+k^2_y\right)
\ee
in other words, no underlying lattice is considered here.

\begin{figure}
\includegraphics[width=0.45\textwidth]{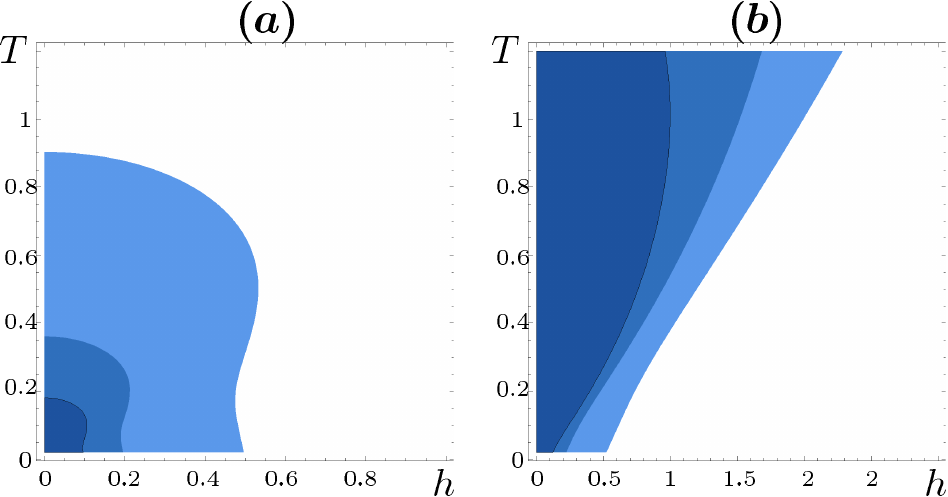}
\caption{(Color online) Instability regions (shaded) for different fillings (from light to dark coloured lines) $\mu=0.1, 0.2, 0.5$ and interaction strength $2\pi mU=0.1$ in fig(a) and $2\pi m U=0.15$ in fig(b). The boundaries of such regions indicate the critical temperature as a function of the external magnetic field.}
\label{fig:nolattice}
\end{figure}

In the case of the spin-antisymmetric interaction given by eqs. (\ref{eq:intspin3})-(\ref{eq:intspin}), the integrals in the $\langle d|d\rangle$ and $(d|d)$ brackets are easily evaluated by turning into polar coordinates in the $k_x, k_y$ plane. We get
\ba
(d|d)&=& 2\pi m \sum_\alpha\left( F[\mu+\alpha h]-(\mu+\alpha h)F'[\mu+\alpha h]\right)\nonumber
\\
\langle d|d\rangle&=& 2\pi m \sum_\alpha F[\mu+\alpha h]
\ea
from which the stability condition $M>0$ can be easily obtained
\small
\be
%\tilde
2\pi m U\!\! \left[\sum_\alpha\!\left( F[\mu\!+\!\alpha h]\!\!-\!(\mu\!+\!\alpha h)F'[\mu\!+\!\alpha h]\right)\right]^2\!\!-\! \sum_\alpha \!F[\mu\!+\!\alpha h]>0
\ee
\normalsize
With this expression, we can plot, {\it e.g.}, the critical magnetic field $h_c$ as a function of $\mu,T$ for any given $U$, as shown in Fig.\ref{fig:nolattice}.

\section{Application to $ Sr_3Ru_2O_7 $}
\label{sec:rutenato}

In this section we apply the present method to the model Hamiltonian\cite{yamase2010}:
\small
\be
\label{hamiltyamase}
H \!=\! \sum_{\mathbf{k},\alpha}\! \left(
( \varepsilon^o(\vk) \!-\! \mu  \!-\! \alpha h) n_{\alpha}\!(\vk)
\!+\!
\frac{1}{2N} \!
\sum_{\vk',\beta} \!f^{\alpha\beta}\!(\mathbf{k},\mathbf{k'}) n_{\alpha}\!(\vk) n_{\beta}\!(\vk) \right)
\ee
\normalsize
where the dispersion relation is that of a $2$-dimensional square lattice with first and second neighbors hopping.
\be
\varepsilon^o(\vk)= - 2 \left(t( \cos{k_x} +\cos{k_y})+ 2 t' \cos{k_x} \cos{k_y}\right)
\ee
The interaction function takes into account only forward scattering and has the form
\be\label{interac}
 f^{\alpha\beta}(\mathbf{k},\mathbf{k'}) = -U e^{\alpha\beta}
 %\delta_{{\alpha\beta}}
 \left(\cos{k_x} - \cos{k_y}\right)\left(\cos{k_x'} - \cos{k_y'}
 \right),
\ee
where $e^{\alpha\beta}$ is a matrix with all its entries equal to $1$.
As discussed  in section \ref{sec:method}, this interaction is separable with the function $d(\vk)$ taking the form of a $d$-wave form factor
\be
d^\alpha(\mathbf{k})= e^\alpha(\cos{k_x} - \cos{k_y})
\label{eq:dwavespint}
\ee
where $e^{\alpha}$ is a vector with all its entries equal to $1$.

\subsection{Mean field calculation}
Within the mean field approximation, the free energy reads\cite{quintanilla}:
 \ba\label{freeenergy}
 F &\simeq&
\sum_{\vk,\alpha}
 \left(
 \varepsilon^o(\vk) -\mu - \alpha h-
 \varepsilon_{\alpha}(\vk)
 \right)
 n_{\alpha}(\vk)
 +
 \nonumber\\&&+\,
 \frac{1}{2N} \sum_{\vk,\vk',\alpha,\beta} f^{\alpha\beta}(\vk,\vk') %(1-\delta_{\bold{k},\bold{k'}}\delta_{\sigma,\sigma'})
  n_{\vk\alpha} n_{\vk'\beta}-
 \nonumber\\&&-\,
 k_BT \sum_{\vk,\alpha} \ln\left(1+ e^{-\varepsilon_{\alpha}(\vk)/k_BT}\right).
\ea
By minimizing it, we find the self-consitency equation
\be\label{eqimplicita}
 \varepsilon_{\alpha}(\vk)= \varepsilon^o(\vk) -\mu - \alpha h-\frac{  1}{N}d^\alpha(\vk) \eta.
\ee
where use has been made of eqs. (\ref{interac}) and (\ref{eq:dwavespint}). In this expression we have defined
\be \label{eta}
\eta= U \sum_{\vk,\beta} \frac{d^\beta(\vk) }{ 1 + e^{\varepsilon_{\beta}(\vk)/k_BT} },
\ee
For each set of parameters $\{ h,\,T,\,\mu,\,U,\,t,\,t' \}$,
$\eta$ is a real number to be found self consistently. Indeed, by writing explicitly the dispersion relation $\varepsilon^o_\alpha(\vk)$ and the $d$-wave form factor $d^\alpha(\vk)$ in expression (\ref{eta}), we get
\ba
\label{eqimplicita2}
\varepsilon_{ \alpha}(\mathbf{k})&=& -\left(2 t +\frac{\eta}{N} \right)\cos{k_x} - \left(2 t - \frac{\eta}{N}\right)\cos{k_y} \nonumber \\
&&+ 4 t' \cos{k_x} \cos{k_y} -\mu - \alpha h,
\ea
Here we see that whenever $\eta$ vanishes, the system has $\pi/4$ rotational invariance. On the other hand whenever $\eta$ is different from zero, isotropy is broken and we have a nematic phase \cite{yamase2005}. For that reason, we will call $\eta$ our ``order parameter'' from now on.

\vspace{.3cm}

First of all, notice that the solution $\eta=0$ 
(the isotropic Fermi liquid solution) is consistent at any point of the parameter space.

%Indeed, by putting $\eta=0$ in eq. (\ref{eqimplicita}) we see 
%that $\varepsilon_{\alpha}(\vk)$ is symmetric in $k_x$, $k_y$. This makes 
%the denominator inside the sum in eq. (\ref{eta}) symmetric. Being the numerator $d^\alpha(\vk)$ 
%antisymmetric in the same variables, this implies that the sum vanishes, making thus 
%evident the consistency of the $\eta=0$ solution.
%Then we have an isotropic Fermi liquid solution at any point of the parameter space.

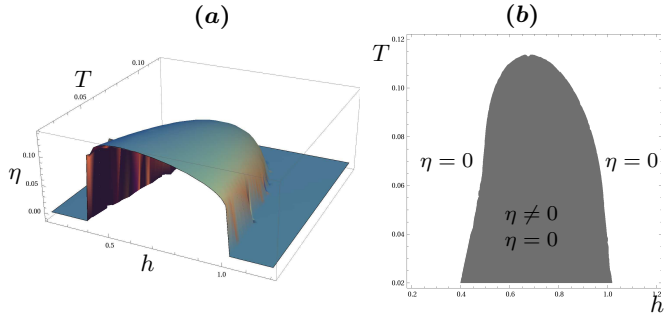
\begin{figure}
\includegraphics[width=0.25\textwidth]{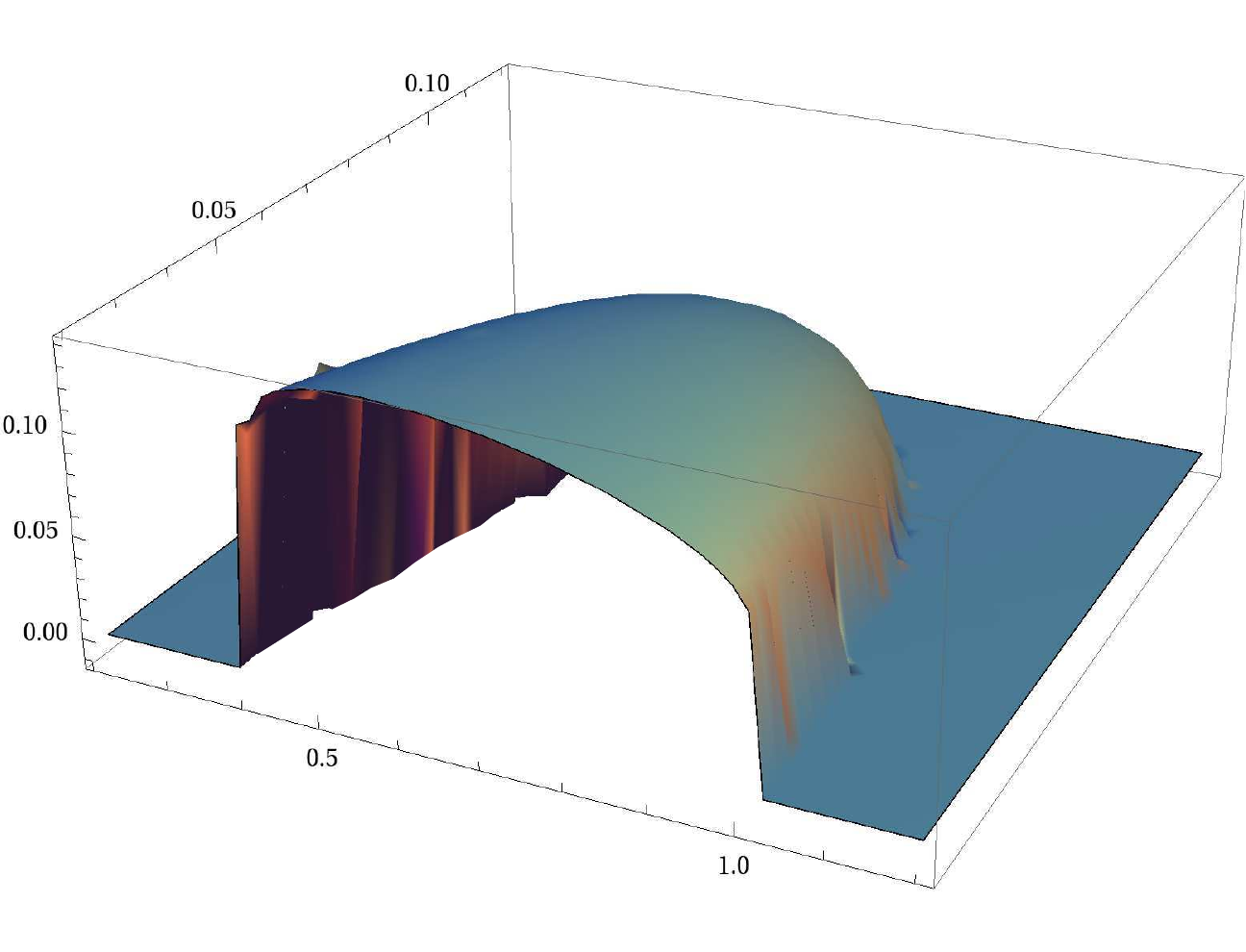}
\hspace{2mm}
\includegraphics[width=0.2\textwidth]{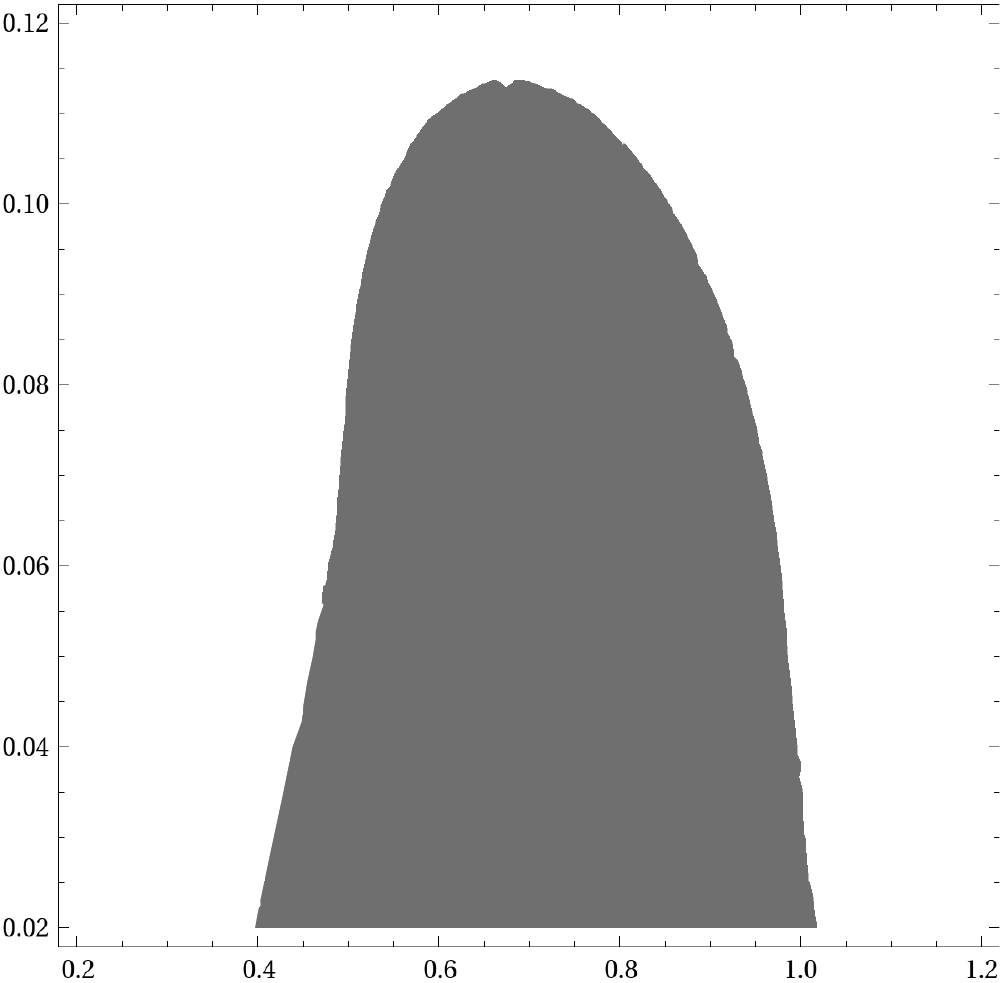}
\begin{picture}(1,1)(0,0)
\put(-180,104){{\boldmath $(a)$}}
\put(-200,10){$h$}
\put(-225,80){$T$}
\put(-250,45){$\eta$}
\end{picture}
\begin{picture}(1,1)(0,0)
\put(60,115){{\boldmath $(b)$}}
\put(115,5){$h$}
\put(10,100){$T$}
\put(28,60){{\footnotesize$\eta= 0$}}
\put(60,40){{\footnotesize $\eta\neq 0$ }}
\put(60,30){{\footnotesize$\eta= 0$}}
\put(98,60){{\footnotesize $\eta= 0$}}
\end{picture}
\caption{(Color online)(a)Order parameter as a function of magnetic field $h$ and temperature $T$. (b) Result obtained for $U=1$ and $\mu=1$. The solution with $\eta =0$ exist at any point of the parameter's space. On the other hand, the $\eta\neq 0$ solution only exists inside the grey dome-shaped region. }
\label{orderparam}
\end{figure}

In addition, one also finds numerically the existence of solutions of eqs.(\ref{eqimplicita})-(\ref{eta}) with $\eta \neq 0$. We have explored the $h$ {\em vs.} $T$ plane for $t=1$, $t'=0.35t$, $U=t$, $\mu=1$, and the results obtained for the order parameter $\eta$ as a function of temperature and magnetic field are shown in Fig \ref{orderparam}(a). As expected, these results coincide with those presented in Ref. [\onlinecite{yamase2005}], implying the presence of a nematic Fermi liquid phase inside a bounded region of the $h$ {\em vs.} $T$ plane. This region is shown in Fig.\ref{orderparam}(b).

\subsection{Extended Pomeranchuk analysis}

With the numerical values obtained for the order parameter $\eta$ inserted into the dispersion relation, we can use our extended Pomeranchuk method to check the stability of the isotropic and nematic phases at any given point of the parameter space at which they are defined.

The isotropic phase is a solution of the self consistency equations at any point of the parameter space, and using our method we find that it is unstable in the region depicted in Fig. \ref{inest}(a). On the other hand, the nematic phase is defined in the grey region of Fig. \ref{orderparam}(b) and we find that it is unstable at the contour of the dome as is indicated in Fig. \ref{inest}(b).

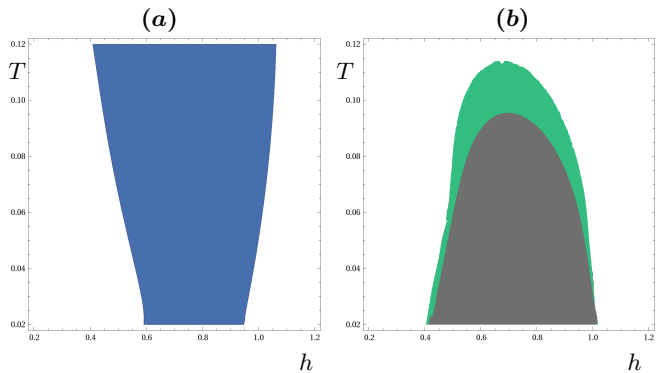
\begin{figure}
\includegraphics[width=0.23\textwidth]{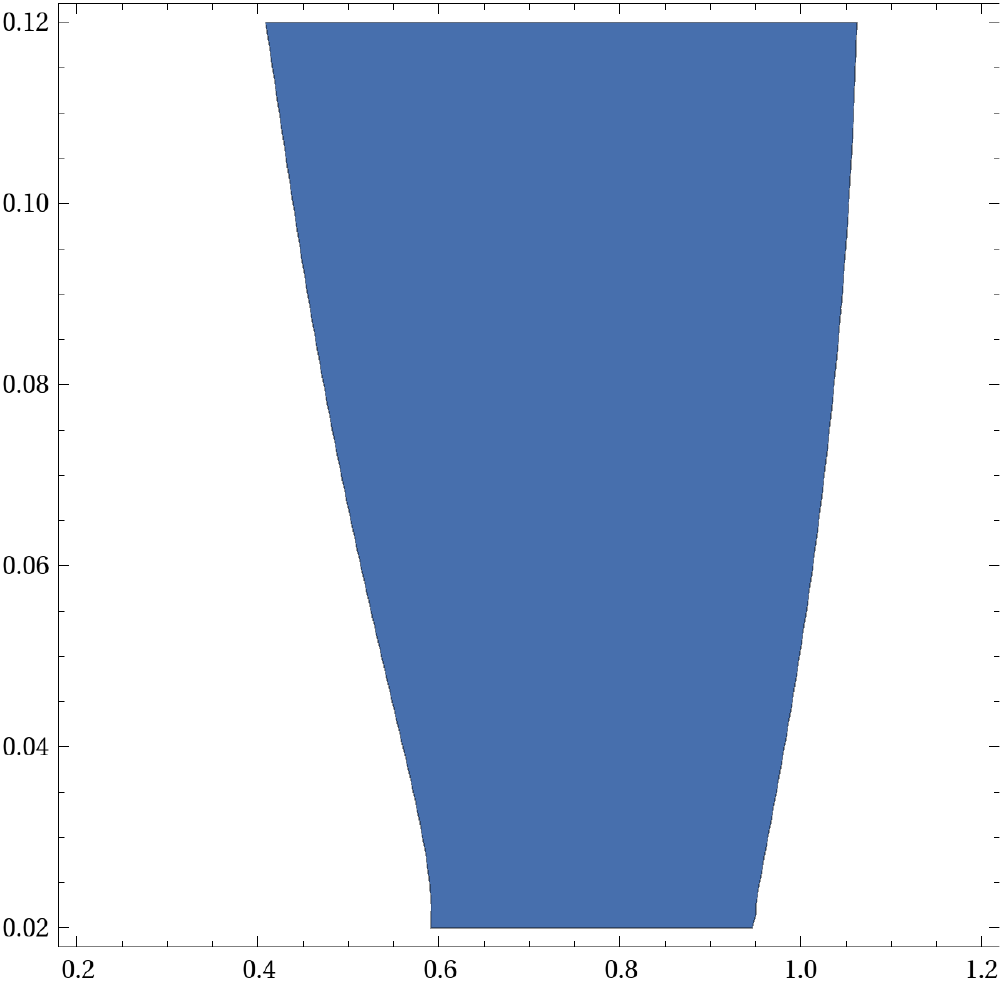}
\hspace{1mm}
\includegraphics[width=0.23\textwidth]{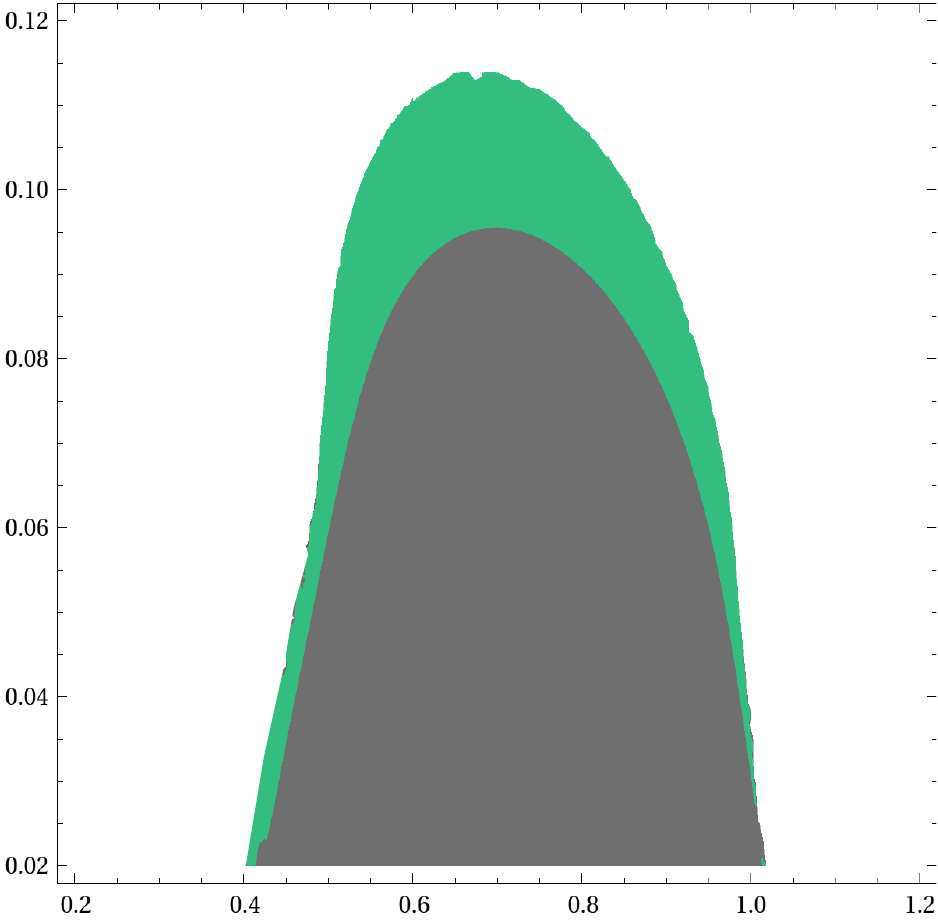}
\begin{picture}(1,1)(0,0)
\put(-70,130){{\boldmath $(a)$}}
\put(-10,0){$h$}
\put(-120,110){$T$}
\end{picture}
\begin{picture}(1,1)(0,0)
\put(60,130){{\boldmath $(b)$}}
\put(110,0){$h$}
\put(0,110){$T$}
\end{picture}
\caption{(Color online) (a) According to the mean field analysis the isotropic phase ({\em i.e.} $\eta =0$) is defined in all the $T$ {\em vs.} $h$ plane. The instability check performed with the extended Pomeranchuk method, shows that it is unstable in the blue region and stable in the rest of the plane. (b) The nematic phase ({\em i.e.} $\eta \neq 0$) is defined only inside the dome shaped region. According to the extended Pomeranchuk method, it becomes unstable in the green region.}
\label{inest}
\end{figure}
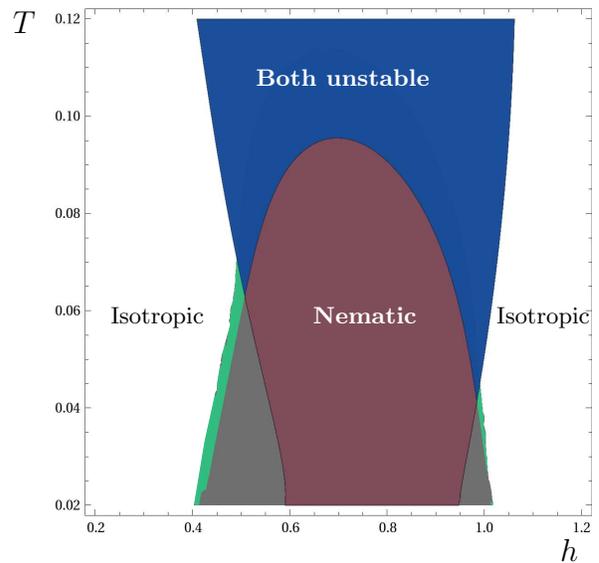
\begin{figure}
\includegraphics[width=0.4\textwidth]{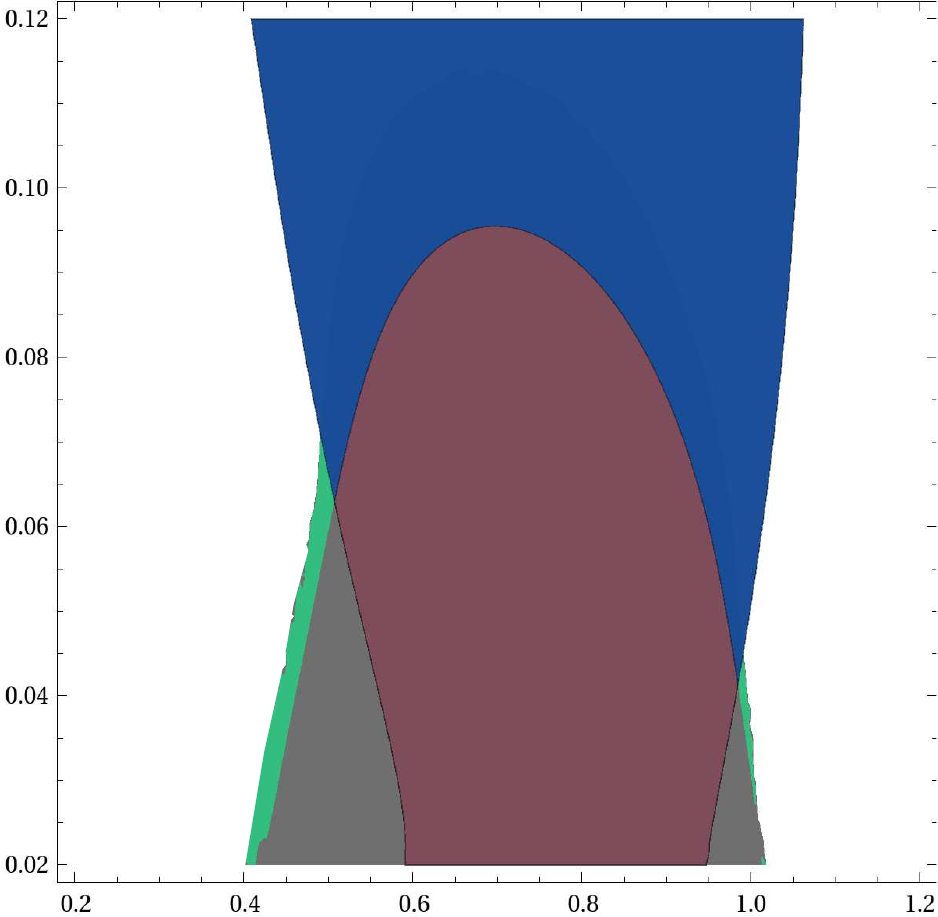}
\begin{picture}(1,1)(0,0)
\put(-15,-10){\large $h$}
\put(-222,190){\large $T$}
\put(-185,80){Isotropic}
\put(-39,80){Isotropic}
\put(-115,80){ \color{blanco}\textbf{ Nematic } }
\put(-130,170){\color{blanco} \textbf{Both unstable}}
\end{picture}
\caption{(Color online.) Phase diagram: In the white and green regions, only the isotropic Fermi liquid phase is stable while in the red region the only stable phase is the nematic Fermi liquid. Inside the grey regions both phases are stable and finally, in the blue one, neither of both Fermi liquids are stable.}
\label{tutti}
\end{figure}
\vspace{1cm}
We can now put together these results to construct the phase diagram shown in Fig.\ref{tutti}, which is the main result of this section. There we can distinguish the two different phases: the isotropic Fermi liquid phase, corresponding to white and green regions and the nematic Fermi liquid phase in the red region. Remarkably, by increasing the temperature it is possible to reach the blue region, where both (isotropic and nematic) Fermi liquids are unstable, indicating the breakdown of the Fermi liquid description. In this way, by complementing the mean-field analysis with the extended Pomeranchuk analysis, the non-Fermi liquid region of the phase diagram is obtained.

Yet another interesting feature of the obtained phase diagram is the existence of regions where both phases are predicted to be stable. These regions are indicated in gray in Fig. \ref{tutti}. As our method could only predict instabilities associated with continuous phase transitions, wherever we detect a possible coexistence of different phases, we must compare their free energies in order to identify the actual phase, that would be reached via a discontinuous transition, as discussed in Ref. \onlinecite{yamase2005}.

\subsection{Discontinuous phase transitions}

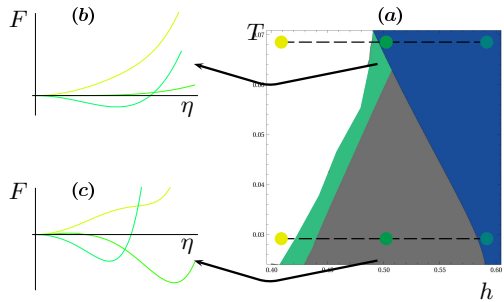
\begin{figure}
\vspace*{0.5cm}
\includegraphics[width=0.35\textwidth]{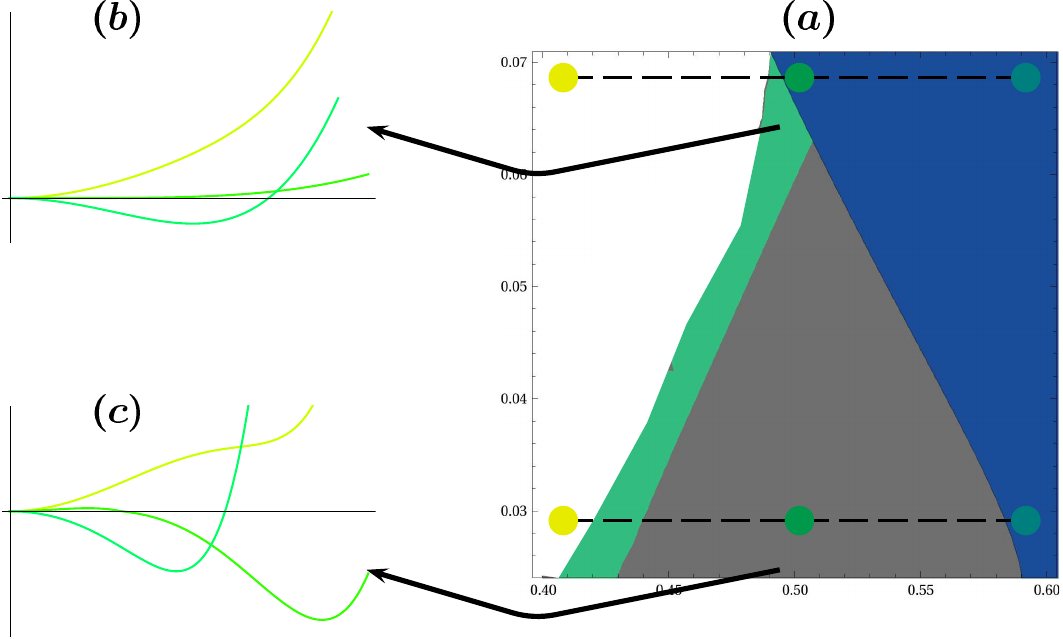}
\begin{picture}(1,1)(0,0)
\put(-12,-3){$h$}
\put(-100,95){$T$}
\put(-125,68){$\eta$}
\put(-125,16){$\eta$}
\put(-190,100){$F$}
\put(-190,35){$F$}
\end{picture}
\caption{(Color online) (a) Zoom of the phase diagram showed in Fig.\ref{tutti}, near one of the coexistence regions. (b) Free energy as a function of the order parameter $\eta$, at T=0.08. There we see a continuous phase transition. When the temperature is lowered we have a discontinuous phase transition, as is shown in (c).}
\label{transitions}
\end{figure}

%A priori, the curvature of the boundary between both Fermi liquid phases could go from concave to convex according which of the phases is the favoured. As the boundary curvature is related with the difference of entropy between the phases by clausius-clapeyron relation, and particularly $Sr_3Ru_2O_7$ exhibit an excess of entropy on the nematic phase \cite{grigeraentropy}, we explored further this issue.
%
In the presented phase diagram, we still have to determine which of the phases survive in the coexistence regions. To do that, we will use the mean field expression for the free energy, eq. (\ref{freeenergy}), replacing the ansatz (\ref{eqimplicita}) for $\varepsilon_{\alpha}(\vk)$ and using the $d$-wave interaction expression (\ref{interac}). We get
\be
F \simeq \frac{1}{2 N} \eta^2 -\frac{1}{\beta}  \sum_{\vk,\alpha} {\rm ln}[1+e^{-\beta \varepsilon_\alpha(\vk)}].
\ee
We can now expand this free energy as a power series in the order parameter, obtaining
\be \label{Fexpansion}
F(\eta)\simeq F(0)+\frac{1}{2}a_2 \eta^2 + \frac{1}{4!}a_4 \eta^4 +...
\ee
Now with the help of the coefficients $a_2$, $a_4$, etc. we can plot the free energy in the coexistence region. The results are shown in Fig.\ref{transitions}. We see that at low temperatures we have a discontinuous phase transition while, when temperature is increased, the transition becomes continuous. We also note that in the coexistence region the minimum of the free energy is achieved for $\eta \neq 0$, implying that the nematic phase is the one that prevails.

\vspace{.3cm}

In this section we have shown that the extended Pomeranchuk method is a useful tool to complement the mean-field description of the low temperature phase diagram of $Sr_3Ru_2O_7$, reproducing both the isotropic and the nematic FL phases, as well as the non Fermi-liquid phase.

\section{Conclusions}
\label{sec:conclusions}

To summarize, in this paper we have presented a more efficient method to diagnose FS instabilities. Within this approach, by parameterizing appropriately the Landau interaction functions, FS instabilities are detected in a much more economic way as compared with previous approaches. We can now obtain the unstable regions analytically in most of the cases, and in those cases in which numerical evaluation is unavoidable, the numerical effort needed is minimal. In addition no extrapolation to  higher channels is needed to determine the phase boundaries. More importantly, the present method allows to study finite temperature problems without resorting to a low temperature expansion.

We have applied the method to previously studied cases and recovered all known results, while instability conditions are obtained as analytic expressions of the parameters involved. In particular, for a continuum model at finite temperature in an external magnetic field we have found
an implicit relation for the critical temperature as a function of the applied field and the chemical potential.\\

Finally, we have applied our method to a $d$-wave interaction function in the square lattice, which has been proposed to describe nematic Fermi liquids \cite{yamase2010,nematics,grigera2011,grigeraotrosa,*grigeraotrosb,*grigeraotrosc}. By combining a mean-field analysis with the Pomeranchuk stability analysis developed in the present paper we reproduced many features of the experimentally observed phase diagram \cite{grigera2011}.

%\bibliographystyle{phjcp}
%\bibliography{biblio}

%

\end{document}